\documentclass[modern]{aastex61}
\usepackage{graphicx}
\usepackage{amssymb}
\usepackage{amsmath}
\usepackage{epstopdf}
\usepackage{hyperref}

\shorttitle{Estimating Spectra From Photometry}
\shortauthors{Kalmbach \& Connolly}

\begin{document}

%\author{Bryce Kalmbach}
%\date{}
%\maketitle
\title{Estimating Spectra from Photometry}

\correspondingauthor{J. Bryce Kalmbach}
\email{brycek@uw.edu}

\author{J. Bryce Kalmbach}
\affil{Department of Physics, University of Washington, Box 351560, Seattle, WA 98195}
%\email{brycek@uw.edu}

\author{Andrew J. Connolly}
\affil{Department of Astronomy, University of Washington, Box 351580, Seattle, WA 98195}
%\email{ajc@astro.washington.edu}

\begin{abstract}
Measuring the physical properties of galaxies such as redshift frequently requires the use of Spectral Energy Distributions (SEDs). SED template sets are, however, often small in number and cover limited portions of photometric color space. Here we present a new method to estimate SEDs as a function of color from a small training set of template SEDs. We first cover the mathematical background behind the technique before demonstrating our ability to reconstruct spectra based upon colors and then compare to other common interpolation and extrapolation methods. When the photometric filters and spectra overlap we show reduction of error in the estimated spectra of over 65\% compared to the more commonly used techniques. We also show an expansion of the method to wavelengths beyond the range of the photometric filters. Finally, we demonstrate the usefulness of our technique by generating 50 additional SED templates from an original set of 10 and applying the new set to photometric redshift estimation. We are able to reduce the photometric redshifts standard deviation by at least 22.0\% and the outlier rejected bias by over 86.2\% compared to original set for z $\leq$ 3.
\end{abstract}


\section{Introduction}
Modeling Spectral Energy Distributions (SEDs) is at the heart of many methods used to study the properties of galaxies. One example is the use of SEDs to estimate galaxy redshifts from photometry rather than spectra \citep{photoz, templatePhotoz}. Photometric redshift estimation works by establishing a relationship between the colors of galaxies observed in a limited set of filters to the colors derived from a full SED of known redshift. Finding the right SED to compare to the catalog colors can, however, be a challenge. Empirical spectra are limited and sparsely sample the color space of a galaxy catalog while synthetic spectra like those of Bruzual \& Charlot \citep{bc03} are calculated with a discrete set of parameters. This means that photometric redshift estimation requires matching a continuous distribution of galaxy colors to a finite set of colors from the template SEDs. When there are large gaps between template colors this leads to uncertainty and erroneous matching can lead to catastrophic outliers in the redshift estimates of a catalog. It would therefore be useful to have an interpolation scheme that, given an arbitrary set of color values, could produce a realistic SED corresponding to those colors. Linear interpolation is sometimes used to expand SED sets for photometric redshift estimation as in \citet{gorecki}, but this method is not guaranteed to produce realistic SEDs.

Another use for galaxy SEDs is the generation of mock catalogs from galaxy modeling codes. When these codes output a mock catalog they use a set of synthetic SEDs to estimate the colors of the simulated galaxies, but are only able to output SEDs using a grid of values for physical properties (e.g., temperature, metallicity, age) that should realistically be continuous. The ability to create a smooth and reasonable interpolation of SEDs could lead to more realistic color statistics for these mock catalogs.

In this paper, we introduce a method for SED interpolation that aims at realistically producing SEDs across continuous regions of color space. We will show that our method reproduces the colors of SEDs better than other interpolation methods such as nearest neighbor or linear interpolation. In Sections \ref{pca_section} and \ref{gpSection} we explain Principal Component Analysis and Gaussian processes and how we apply them to generate an interpolation scheme in color space for SEDs. In Section \ref{testSection} we present a demonstration of our technique where we generate SEDs at given points in color space and compare the results to other interpolation methods. In Section \ref{photoZ} we will use photometric redshift estimation as an example of the benefits of our method. Finally, we discuss our work in Section \ref{discussion} and conclude in Section \ref{conclusion}.

\section{Creation of basis using Principal Component Analysis} \label{pca_section}

SED fitting techniques attempt to map observed galaxy colors to intrinsic galaxy properties. For instance, photometric redshift estimation attempts to match galaxy colors to their redshifts. As we will show, the sparseness of the template set and its incomplete coverage in color space can create errors in redshift estimation. In order to improve SED fitting techniques we therefore need to be able to interpolate and extrapolate in color space to create a continuous mapping between colors and SEDs. To begin solving this problem, we choose to create a new basis in Section \ref{training_set} from the template SEDs using Principal Component Analysis (PCA). Linear combinations of these new basis SEDs can then be used to construct galaxy SEDs where the coefficients that weight each basis spectrum will be estimated from the color space.

\subsection{Principal component analysis}

PCA, also known as the Karhunen-Lo\`eve transform, has been used on SEDs in astronomy mainly for classification by using a small number of coefficients to describe each spectrum \citep{astroPCA, galaxyClassPCA}. This method identifies the directions of maximal variance in a dataset and in so doing provides a new basis to represent the data along these directions. For our case we call these new bases eigenspectra following the convention of \citet{astroPCA}. Using a linear combination of all these eigenspectra with proper weighting we can reconstruct the original spectra:
\begin{equation} \label{linearCombPCA}
f_{\lambda i} = \sum^{m}_{j=1} y_{ij}e_{\lambda j}
\end{equation}
where $f_{\lambda i}$ is the flux of the $i$th original spectrum at wavelength $\lambda$, $m$ is the total number of eigenspectra and $y_{ij}$ is the coefficient that is applied to $j$th eigenspectrum when reconstructing the $i$th spectrum. If we keep all $m$ eigenspectra we can reconstruct the input model spectra perfectly, but in PCA the eigenvectors are ordered by the variance they describe in the original dataset. If desired, we can truncate the sum in Equation \ref{linearCombPCA} at a value $n < m$ where $n$ is the number of components that reproduces the original spectra within some error tolerance.

\section{Gaussian process Regression for Eigencofficients} \label{gpSection}

Our goal is now to estimate SEDs for points in color space where we want to expand our template set, i.e. generate additional SEDs where the templates do not cover the catalog color space. To do this we want to estimate new PCA coefficients using a regression of the PCA coefficients on to template SEDs based upon their locations in color space. We use Gaussian processes to do this regression because they perform well in creating smooth, nonlinear functions and are able to handle interpolation and extrapolation. \added{Gaussian Processes have entered into astronomy recently in areas such as photometric redshift estimation \citep{gpphotoz_sdss, gpz, gpphotoz} and analysis of time series data \citep{gp_kepler}.} We use maximum likelihood estimation to quickly optimize a new regression function for each of the PCA coefficients since the hyperparameters that estimate each coefficient are not expected to be the same.

\subsection{Gaussian processes}

A Gaussian process (GP) is a continuous distribution of random variables where any finite number of points can be described with a joint Gaussian distribution \citep{gpBook}. For example, consider the 2-d data vector $\mathbf{y(x_{i})} = (y({x_{1}}), ...., y({x_{n}}))$ which is a single draw of a multivariate Gaussian distribution of dimension n. Since $\mathbf{y}$ is the result of a joint Gaussian distribution at points $x_{1},...,x_{n}$ it is also the result of a GP that can be sampled continuously along the x axis \citep{GPintro}. GPs are described by a mean function, $m(\mathbf{x})$, and a covariance function, $k(\mathbf{x}, \mathbf{x'})$, so that a GP can be abbreviated as $f(\mathbf{x}) \sim \mathcal{GP}(m(\mathbf{x}), k(\mathbf{x}, \mathbf{x'}))$ \citep{gpBook}. In general and in our use of GPs, the mean function is assumed to be zero so that the GP can be completely specified by the covariance function that relates data points to one another. Putting this together we can use a GP associated with an observed dataset $\mathbf{y} \sim \mathcal{N}(0, k(\mathbf{x}, \mathbf{x'}))$ to make predictions about the values of other points in the same space as $\mathbf{y}(\mathbf{x})$. This is the procedure known as Gaussian process regression and is better known in some fields as kriging. If the data are noisy we can follow \citet{GPintro} and add in a Gaussian noise component along with the covariance function giving 
\begin{equation} \label{covNoise}
\centering
K(\mathbf{x},\mathbf{x'}) = k(\mathbf{x},\mathbf{x'}) + \sigma_{n}^{2}\delta(\mathbf{x},\mathbf{x'})
\end{equation} 
where $\delta(x,x')$ is the Kronecker delta. Here we assume that our noise is Gaussian and independent, but if there is covariance in the noise we could include additional terms in the covariance function with the desired form \citep{gpBook}. Thus, our problem that we show in practice in Section \ref{estimate_seds} can now be described with our observed data sample $\mathbf{y}$ (in this case the PCA coefficients) and a test sample $\mathbf{y_{*}}$ (our interpolated coefficients) as:
\begin{equation}
\begin{bmatrix}
	\mathbf{y} \\
	\mathbf{y_{*}}
\end{bmatrix}
\sim \mathcal{N}(0, 
\begin{bmatrix}
	K \ K_{*}^{T} \\
	K_{*} \  K_{**}
\end{bmatrix})
\end{equation}
where $K$ is the covariance matrix of the data, $K_{*}$ is a matrix containing the covariances of the data and the test points and $K_{**}$ is the covariance matrix between the test points.

Since we are interested in using this for regression we want to find the conditional probability of a set of test points given our observations, $p(\mathbf{y_{*}}|\mathbf{y})$. Using the equations from Appendices 2 and 3 in \citet{gpBook} this turns out to be another Gaussian distribution:
\begin{equation}
\mathbf{y_{*}}|\mathbf{y} \sim \mathcal{N}(K_{*}K^{-1}\mathbf{y}, K_{**} - K_{*}K^{-1}K_{*}^{T})
\end{equation}
The mean of the new Gaussian at $\mathbf{x_{*}}$ is the best estimate for the value of $\mathbf{y_{*}}$ while the variance of the new Gaussian at that point provides a measure of the uncertainty in the predicted $\mathbf{y_{*}}$.

\subsection{Choice of Kernels and Hyperparameters}\label{kernels}

We do not assume that the coefficients for each principal component will have the same relationships in color space and as a result we train a separate GP for the eigencoefficients of each principal component. The GPs use the color coordinates of the training spectra as inputs $\mathbf{x}$ and the PCA eigencoefficients as the outputs $y(\mathbf{x})$. In section \ref{pca_section} we showed how the eigencoefficients can be used to reconstruct a spectrum using the PCA eigenspectra. Since our goal is to create new SEDs at specific points in color space we use GP regression to estimate new eigencofficients for points in color space. Then we use the coefficients along with the eigenspectra to calculate a new SED. Since the covariances in color space may be different for each of the principal components we use a separate GP regression for each eigencoefficient. We also tune each GP to find the best hyperparameters for the covariance function. These hyperparameters are the terms in each covariance function that affect the strength of the relationship of the points in the training set to a desired measurement location. For instance, the commonly used squared exponential covariance function uses a scale length to adjust the weighting in the GP based upon Euclidean distance to the training points. In order to find the hyperparameters for a set of data, we  maximize the log marginal likelihood function as found in Chapter 5 of \citet{gpBook}:
\begin{equation} \label{eq: marginal likelihood}
\log \ p(\mathbf{y}|X, \theta) = -\frac{1}{2}\mathbf{y}^{T}K^{-1}\mathbf{y} - \frac{1}{2} \log \begin{vmatrix} K \end{vmatrix} - \frac{n}{2} \log 2\pi
\end{equation}
and use K to mean the same input data covariance matrix as above while $n$ refers to the size of the training set. 

We implement our GP methods using the Python language package, $george$\footnote{https://github.com/dfm/george} \citep{george} and use for comparison 4 different kernel functions that come in $george$ to describe our covariance. All are stationary kernels in which the kernel function does not depend on the values of the input coordinates, but only on the distance between points. \replaced{These are the squared exponential, $\theta_{1}\exp(\frac{-d_{i}^{2}}{2\*\theta_{2}})$ and the exponential function, $\theta_{1}\exp(-\sqrt{\frac{d_{i}^2}{\theta_{2}}})$}{These are the squared exponential, $\theta_{1}\exp(\frac{-d_{i}^{2}}{2\*\theta_{2}})$, the exponential, $\theta_{1}\exp(-\sqrt{\frac{d_{i}^2}{\theta_{2}}})$, the Matern-3/2, $\theta_{1}(1 + \sqrt{\frac{3d_{i}^{2}}{\theta_{2}}})\exp(-\sqrt{\frac{3d_{i}^2}{\theta_{2}}})$ and the Matern-5/2, $\theta_{1}(1 + \sqrt{\frac{5d_{i}^{2}}{\theta_{2}}} + \frac{5d_{i}^2}{3\theta_{2}})\exp(-\sqrt{\frac{5d_{i}^2}{\theta_{2}}})$}. $\theta_{1}$ and $\theta_{2}$ are the tunable hyperparameters of each covariance model, where the first is used to set the signal variance in the outputs and the second adjusts the distance scale on each model. To set the values of $\theta_{1}$ and $\theta_{2}$ for each covariance function we maximize the value of Equation \ref{eq: marginal likelihood} using Nelder-Mead optimization implemented using the Scipy library \citep{scipy}. 

\section{Testing and Results} \label{testSection}
To test our method we performed three sets of 500 runs where the goal was to use only the color information for SEDs we already had and compare the estimated SED to the original. The first test looked at the results where we restricted SED estimation to the optical wavelength range covered by the filters we used for the input colors. The other two tests looked at estimating the SEDs over additional wavelengths above and below the optical range.  In each run, we randomly drew a set of 10 templates from a larger set of synthetic SEDs described below to form a training set. We used these spectra as input to our PCA method to create a set of eigenspectra. We then drew another 50 templates to make up a test set where we calculated the colors and assumed we only had this color information as a starting point for SED estimation. We used these colors as input to the Gaussian process regression trained with our training set eigencofficients and generated new eigencoefficients that we used to reconstruct estimates of the original SEDs. Finally, we compared the resulting SEDs to the actual SEDs for our method. We also compared our results for SED estimation against two other common methods, nearest neighbor estimation and linear interpolation. The results were stored and a new run was started with a new set of training and test SEDs randomly chosen from the SED library.

\subsection{Creation of training set} \label{training_set}

As templates we used SEDs from the LSST simulations \citep{sims} SED library\footnote{We used the version current as of January 2017 which can be downloaded at \url{https://lsst-web.ncsa.illinois.edu/sim-data/sed\_library/seds\_170124.tar.gz.}} which are \citet{bc03} (BC03) model SEDs. The full library of 959 spectra samples 4 different star formation histories and span a range of ages from 1.585 Myr to 12.5 Gyr (using the Padova 1994 isochrones). It also includes 6 different metallicities and uses the \citet{chabrier} IMF as described in the BC03 package. \added{Emission lines are not included in these spectra. The top part of Figure \ref{spec_examples} shows a set of sample SEDs with different ages, metallicities and star formation histories. The spectra cover a wavelength range from 9 $nm$ up to 160 $\mu m$, but we only used wavelengths less than 2400 $nm$ where the resolution is best. Here the resolution varies between 0.1 $nm$ for most of the optical range up to 10 $nm$ at longer wavelengths and values in between for other wavelengths. The bottom part of Figure \ref{spec_examples} shows a more detailed look at the resolution throughout the wavelength range. 
\begin{figure}
	\centering
    	\includegraphics[width=\linewidth]{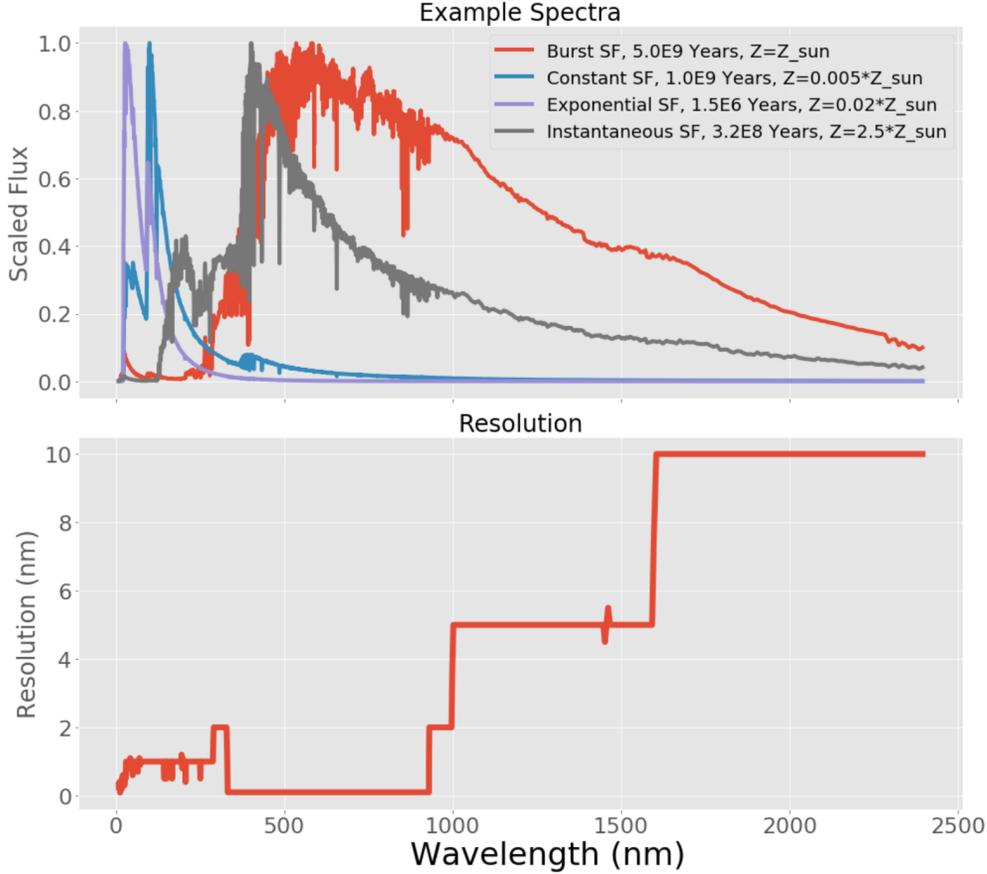}
	\caption{Top: Example spectra from LSST simulations SED library used in this work. Bottom: Resolution of the spectra as a function of wavelength.}	\label{spec_examples}
\end{figure}
} Since some spectra in the library are very similar and the Gaussian processes require non-singular matrices for matrix inversion we trimmed the catalog to 789 spectra by removing spectra that duplicated the flux of another within 0.001\% at more than 90\% of the wavelength points.

\begin{figure}
	\centering
    	\includegraphics[width=\linewidth]{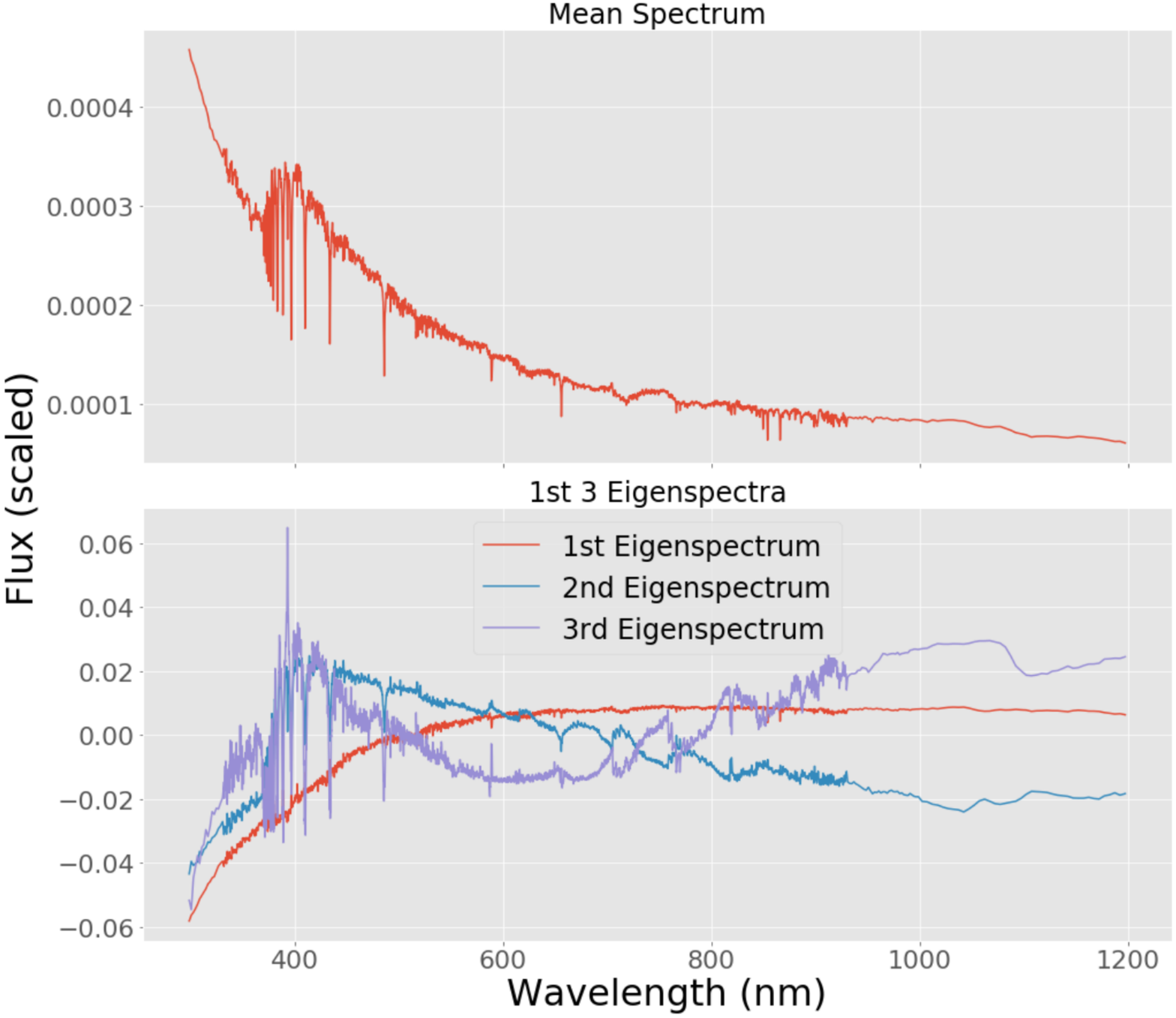}
	\caption{Mean spectrum and first three eigenspectra of the PCA performed on a randomly chosen set of 10 BC03 spectra.}	\label{pca_eigenspectra}
\end{figure}

In each run we created a new basis set from 10 spectra randomly picked from the full set of 789 spectra. We used 10 spectra in order to provide a training set comparable in size to photometric redshift training sets like the commonly used \citet{cww} (CWW) templates. Then we used the PCA method in the decomposition module of scikit-learn \citep{scikit} to find eigenspectra and eigencoefficients. \deleted{While the full spectrum has wavelengths from 9 $nm$ up to 160 $\mu m$ we ran tests with different subsets of these wavelengths.} In each PCA we kept 9 of the 10 components since we found that 9 components in each PCA decomposition explained over 99.9999\% of the variance. We show in Figure \ref{pca_eigenspectra} the mean spectrum and the first three eigenspectra for one of our training sets in \ref{optical_test}.

\subsection{Estimating SEDs} \label{estimate_seds}

In order to estimate new eigencofficients at locations in color space we used the exponential and squared exponential kernels described in \ref{kernels} for our Gaussian processes. Since we implement a separate Gaussian process for each set of eigencoefficients we ended up with 9 different Gaussian processes from each training set. 

We then used the Gaussian processes and the colors of our test set to predict eigencoefficients for the test spectra and used this information along with the eigenspectra and mean spectrum of the training set to generate estimates of the test SEDs. Since we use Principal Component Analysis for our basis set there is a possibility of negative flux at some points in the estimated SEDs. In these cases we set the flux to 0 where it would otherwise be negative.

To generate our linear estimation comparison set we trained a linear regression in color space with the training SEDs and then used this to estimate the flux values at the colors of our test spectra. For the nearest neighbor comparisons we experimented with different variations where the nearest neighbors were determined by distance in color space between the test colors and the nearest training colors. We tested using a uniform weighting with 1, 2 or 4 neighbors as well as a distance weighted estimate using 2 or 4 neighbors.

\subsubsection{Optical Wavelengths} \label{optical_test}
The colors we plan on using for this technique are most likely those of an optical survey such as the Large Synoptic Survey Telescope (LSST). Using the latest version of the LSST bandpasses from the LSST simulations software stack \citep{sims} \footnote{We used the versions current as of March 2017 found here: https://github.com/lsst/throughputs} we calculated the colors for our training and test SEDs. We followed the procedure outlined above only using the SEDs at wavelengths from 299 - 1200 $nm$ which covers the range of the LSST filters for the PCA stage. The mean maximum likelihood hyperparameters across all 500 runs for the first three eigencoefficients are shown for each of the kernel functions in Table \ref{theta_fits} where, as in Section \ref{kernels}, $\theta_{1}$ is a scaling factor and $\theta_{2}$ is a length factor. Notice that the $\theta_{1}$ and $\theta_{2}$ hyperparameters in each model vary with each PCA component as expected.
\begin{deluxetable}{c|cccc}
\tablecaption{Mean Gaussian Process hyperparameter values}
%\begin{table}
%\centering
%\resizebox{\columnwidth}{!}{%
%\begin{tabular}{l|ll}
%\hline
%\multicolumn{1}{c}{\textbf{}} & \multicolumn{1}{c}{\textbf{Exponential}} & \multicolumn{1}{c}{\textbf{Sq. Exponential}}  \\ \hline
\tablehead{
\colhead{Principal Component} & \colhead{Exponential} & \colhead{Squared Exponential} & \colhead{Matern-3/2} & \colhead{Matern-5/2}\\
\colhead{} & \colhead{$(\theta_{1}, \theta_{2})$} & \colhead{$(\theta_{1}, \theta_{2})$} & \colhead{$(\theta_{1}, \theta_{2})$} & \colhead{$(\theta_{1}, \theta_{2})$}
}
\startdata
1st & (5.14E-5, 6.16E+1) & (1.65E-4, 9.66) & (9.21E-4, 2.57E+2) & (2.75E-4, 3.60E+1) \\
2nd & (1.58E-6, 6.09E-1) & (3.66E-5, 2.14) & (6.58E-5, 1.60E+1) & (8.09E-5, 8.94) \\
3rd & (9.94E-8, 1.23E-1) & (9.93E-6, 1.19) & (3.49E-5, 3.70E+1) & (3.39E-5, 5.71) \\
%\end{tabular}%
\enddata
%}
\tablecomments{Mean Gaussian Process hyperparameter values for 1st three principal components after 500 runs. $\theta_{1}$ is a scaling factor and $\theta_{2}$ is a length factor.} \label{theta_fits}
\end{deluxetable}
%\end{table}

%\begin{table}
%\centering
%\resizebox{\columnwidth}{!}{%
\begin{deluxetable}{cc|ccccccc}
\tablecaption{Percentage residual errors in flux in \ref{optical_test}}
%\hline
\tablehead{
\colhead{Error Metric} & \colhead{Wavelength Range} & \colhead{Exp.} & \colhead{Sq. Exp.} & \colhead{Mat.-3/2} & \colhead{Mat.-5/2} & \colhead{NN} & \colhead{2 NN} & \colhead{Linear} \\
\colhead{} & \colhead{nm} & \colhead{} & \colhead{} & \colhead{} & \colhead{} & \colhead{} & \colhead{} & \colhead{}
}
\startdata
Mean Error & 299 - 1200 & 7.25\% & 3.32\% & 3.60\% & 3.17\% & 9.70\% & 9.23\% & 9.62\%  \\
Interquartile Mean Error & 299 - 1200 & 4.04\% & 2.09\% & 2.63\% & 2.11\% & 7.39\% & 6.18\% & 6.65\% \\
\enddata
%\end{tabular}%
%}
\tablecomments{Residuals are errors in flux between true and estimated SEDs.} \label{test_1_results}
\end{deluxetable}

\begin{figure}
 	\centering
        \includegraphics[width=\linewidth]{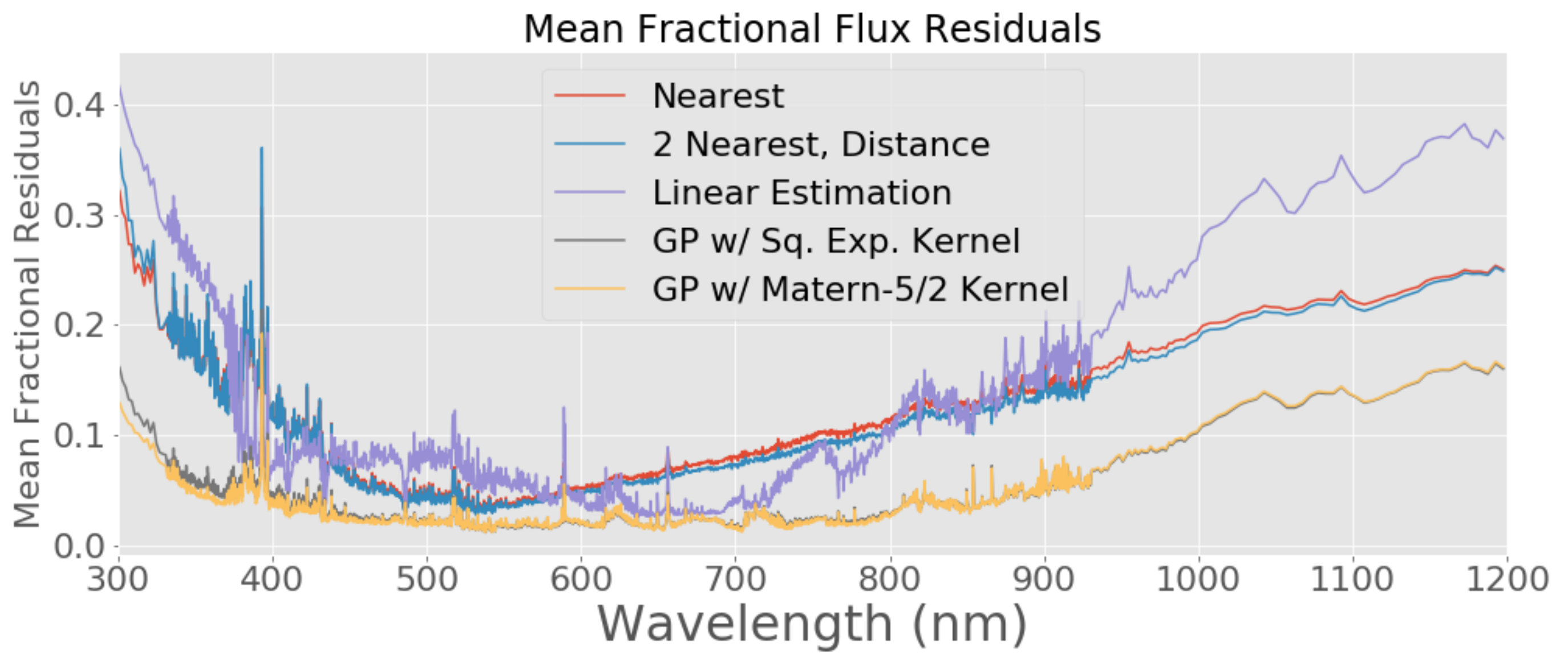}
	\caption{Mean fractional residuals between predicted and original spectra over 500 runs. Gaussian processes with a Matern-5/2 kernel outperformed across almost all wavelengths.}	\label{spec_error}
\end{figure}

\begin{figure}
 	\centering
		\includegraphics[width=\linewidth]{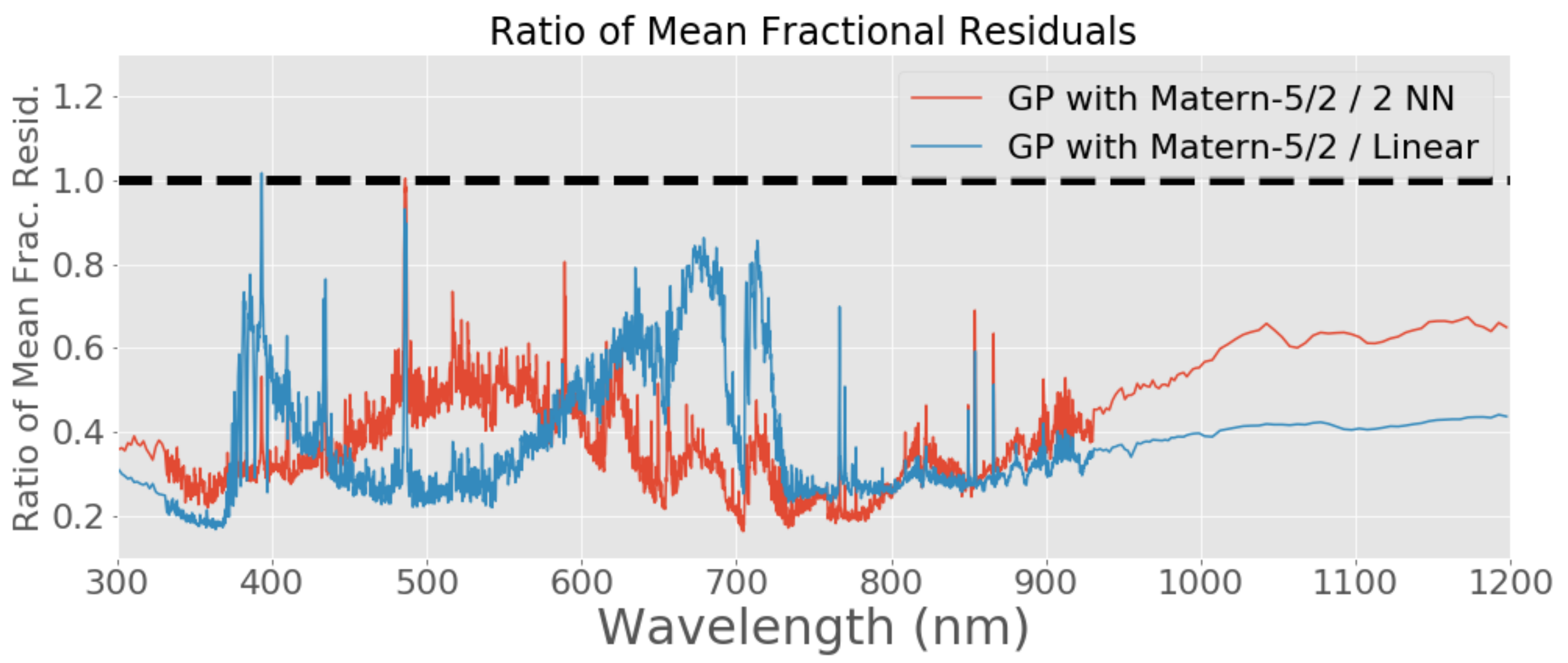}
	\caption{Ratio of our method's fractional residuals to that from 2 nearest neighbors with distance weighting and to that from linear interpolation. The line drawn at 1.0 shows the level at which the errors in each method would be the same. Values below this line mean that the Gaussian Process estimated SEDs have a lower mean residual than those of the alternate method. The Matern-5/2 kernel has less error than any other method at almost every wavelength.}	\label{spec_error_ratio}
\end{figure}

The output of each of the 500 runs is a set of 50 estimated SEDs from each estimation method at the locations in color space of 50 original test SEDs. To compare the results we find the absolute difference between the estimated and original SED for each method and then calculate the fractional residuals. Figure \ref{spec_error} displays the mean residuals between the predicted SEDs and the actual SEDs as a function of wavelength for the two best Gaussian Process estimates and the two best nearest neighbor results along with the linear estimation. The results show that our GP method with a \replaced{squared exponential}{Matern-5/2} kernel outperforms all other techniques across almost all wavelengths that we used. The same technique with a \replaced{exponential}{squared exponential} kernel also works well compared to the alternatives. In Figure \ref{spec_error_ratio} we show the ratios between our method and the comparison methods. The best nearest neighbor technique in this test was the distance weighted 2 nearest neighbors estimate and this is what we include for comparison in Figure \ref{spec_error_ratio}. For most of the spectra our method using the \replaced{squared exponential}{Matern-5/2} kernel has less than 60\% of the error as the nearest neighbor method and less than 50\% of the error as the linear estimation method. The mean error and median error across the spectrum is shown in Table \ref{test_1_results}. Comparing the different methods we see that the mean percent error was 3.20\% and 3.49\% for the Matern-5/2 kernel and squared exponential kernel methods, respectively. This is lower than the 9.70\% mean error for the nearest neighbor method, the 9.23\% for the 2 nearest neighbors method and the 9.62\% mean error when using linear estimation. \added{In fact, all four GP kernels outperformed the nearest neighbor and linear methods.}

\subsubsection{Expanding the wavelength range} \label{test_2}

It is important for applications involving redshifts or predicting magnitudes in other bands to be able to use our technique to generate accurate SEDs at a larger wavelength range than just the rest frame optical wavelengths. Therefore our second test was using the same training and test SEDs with the same colors and performing the same analysis, but extending the wavelengths to 99 and 2400 $nm$ in our training SEDs for use in the creation of the PCA basis.

%\begin{table}
%\centering
%\resizebox{\columnwidth}{!}{%
\begin{deluxetable}{cc|ccccccc}
%\hline
\tablecaption{Percentage Residual Errors in flux for \ref{test_2}}
\tablehead{
\colhead{Error Metric} & \colhead{Wavelength Range} & \colhead{Exp.} & \colhead{Sq. Exp.} & \colhead{Mat.-3/2} & \colhead{Mat.-5/2} & \colhead{NN} & \colhead{2 NN} & \colhead{Linear} \\
\colhead{} & \colhead{nm} & \colhead{} & \colhead{} & \colhead{} & \colhead{} & \colhead{} & \colhead{} & \colhead{}
}
\startdata
Mean Error & 99 - 2400 & 43.7\% & 73.5\% & 51.2\% & 60.1\% & 27.1\%  & 33.6\% & 56.5\% \\
Mean Error & 299 - 1200 & 14.7\% & 19.8\% & 16.1\% & 17.6\% & 14.9\% & 14.2\% & 16.1\% \\
Interquartile Mean Error & 99 - 2400 & 13.4\% & 15.9\% & 15.4\% & 14.6\% & 13.7\% & 13.2\% & 15.9\%  \\
\enddata
%\end{tabular}
%}
\tablecomments{Residuals are errors in flux between true and estimated SEDs.} \label{test_2_error}
\end{deluxetable}

Table \ref{test_2_error} shows that our methods are not performing as well in this test as they did in the first one. The mean error across the full spectrum is now as high as 73.5\% for the squared exponential kernel and even the best GP kernel, the exponential is at 43.7\%.
The traditional methods also show increases in error but now outperform the Gaussian process regression methods. Here the exponential kernel performs better than the other kernels but still is worse than the nearest neighbor and linear methods across most of the spectrum. Looking at the second row of Table \ref{test_2_error} however, we see that the errors in the region of the spectrum used to train the GPs are actually comparable between the Nearest Neighbor method and our GP method. \added{Since all SED sets in this test are the same as those we used in \ref{optical_test} it appears that the addition of wavelengths outside the range of our filters are leading to the decrease in performance of our method with most of the error appearing at the blue end of the spectrum.} We therefore decided to try and find ways to use information about the rest of the spectrum available to our training set in order to improve our estimates for the test set where we only use the LSST filters.

\subsubsection{Using artificial filters in training} \label{training_new_filters}
Since it seems that the errors are driven by features outside the range of our filters we decided to expand the amount of information we were using in our training set. We added a series of top hat filters outside the range of the LSST \textit{ugrizy} filters we were using. We tried a few simple combinations of 50 $nm$ wide top hat filters close to the existing bands and the best results came from 2 50 $nm$ wide top hat filters on the blue end at 100-150 $nm$ and 200-250 $nm$ as well as 2 on the red end at 1250-1300 $nm$ and 1350-1400 $nm$. 

\begin{figure}
 	\centering
		\includegraphics[width=\linewidth]{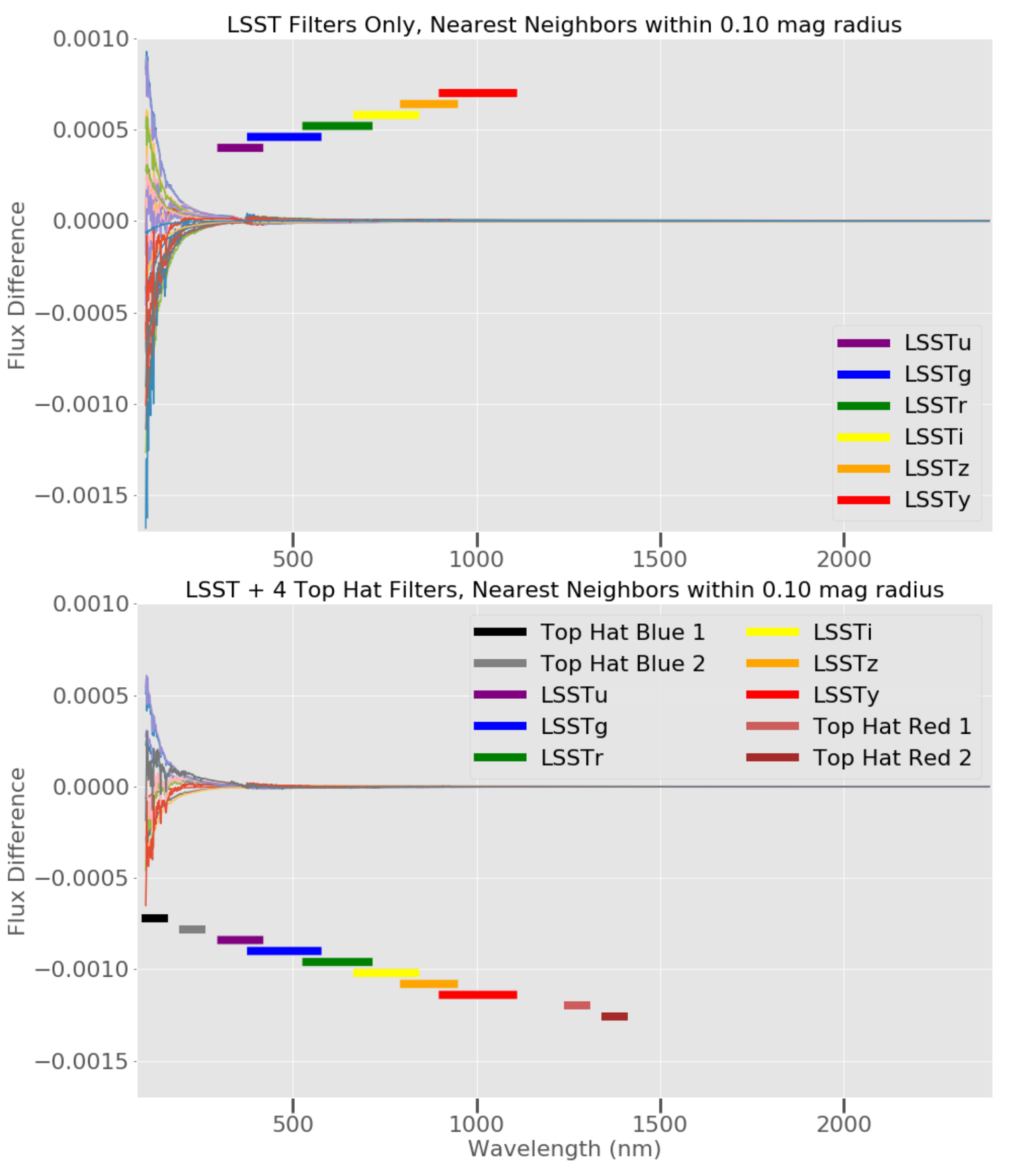}
	\caption{Top: Difference between a single BC03 spectrum and all other spectra within a radius of 0.1 mags in the 5-dimensional LSST color space. The wavelength span of the LSST filters are shown as the colored horizontal bars. Bottom: Difference between a single BC03 spectrum and all other spectra within a radius of 0.1 mags in the 9-dimensional LSST+4 top hat filters color space. The wavelength span of the LSST filters and the added top hat filters are shown as the colored horizontal bars.}	\label{UV_diversity}
\end{figure}

First, we wanted to test the hypothesis that the relationship of the spectra in the UV and IR regions were not being fully captured by only using the optical filters. We took our full set of BC03 spectra and picked out a test spectrum. We then took all of the other spectra within a radius of 0.1 magnitudes in the 5-dimensional color space provided by the LSST filters. The top plot in Figure \ref{UV_diversity} shows the difference in flux between these spectra and the test spectrum. Then we repeated this process with the same spectrum, but in the 9-dimensional color space of the LSST \textit{ugrizy} plus our top hat filters in the UV and IR regions of the spectrum. Once again we used a radius of 0.1 magnitudes and show the results in the bottom plot of Figure \ref{UV_diversity}. Notice that in the comparison of the spectra with similar optical colors the SEDs are very similar across the optical and IR wavelengths. In the UV portion of the SED (wavelengths less than the \textit{u} band filter) there is, however, significant diversity in the spectra. When adding the top hat filters to expand the color space and selecting spectra with common optical and UV colors the large amount of diversity in the UV part of the spectrum is removed
(for spectra within the same 0.1 magnitude radius of our test spectrum). 

Confident that the top hat filters would help to better fit the relationships between spectra in color space we moved on to using the top hat filters in our GP Regression. Since we would not have this information when applying our technique to observed colors we only used this to optimize the hyperparameters for the GP on the training set information. We then used these optimized hyperparameters with the same 5 color GPs as before on the test set data. Therefore, we simulated having full spectra in our templates but only the observed color information we would have in a real application as the data for our test set estimates.

%\begin{table}
%\centering
%\resizebox{\columnwidth}{!}{%
\begin{deluxetable}{cc|cccc}
\tablecaption{Percentage Residual Errors in flux for \ref{training_new_filters}}
%\hline
\tablehead{
\colhead{Error Metric} & \colhead{Wavelength Range} & \colhead{Exp.} & \colhead{Sq. Exp.} & \colhead{Mat.-3/2} & \colhead{Mat.-5/2} \\
\colhead{} & \colhead{nm} & \colhead{} & \colhead{} & \colhead{} & \colhead{}
}
\startdata
Mean Error & 99 - 2400 & 32.1\% & 72.3\% & 25.6\% & 44.7\%  \\ 
Mean Error & 299 - 1200 & 12.1\% & 21.6\%  & 13.1\% & 16.9\%\\ 
Interquartile Mean Error & 99 - 2400 & 11.1\% & 16.8\% & 15.1\% & 14.2\%\\ 
\enddata
%\end{tabular}%
%}
\tablecomments{Residuals are errors in flux between true and estimated SEDs.} \label{test_3_error}
\end{deluxetable}

\begin{figure}
 	\centering
		\includegraphics[width=\linewidth]{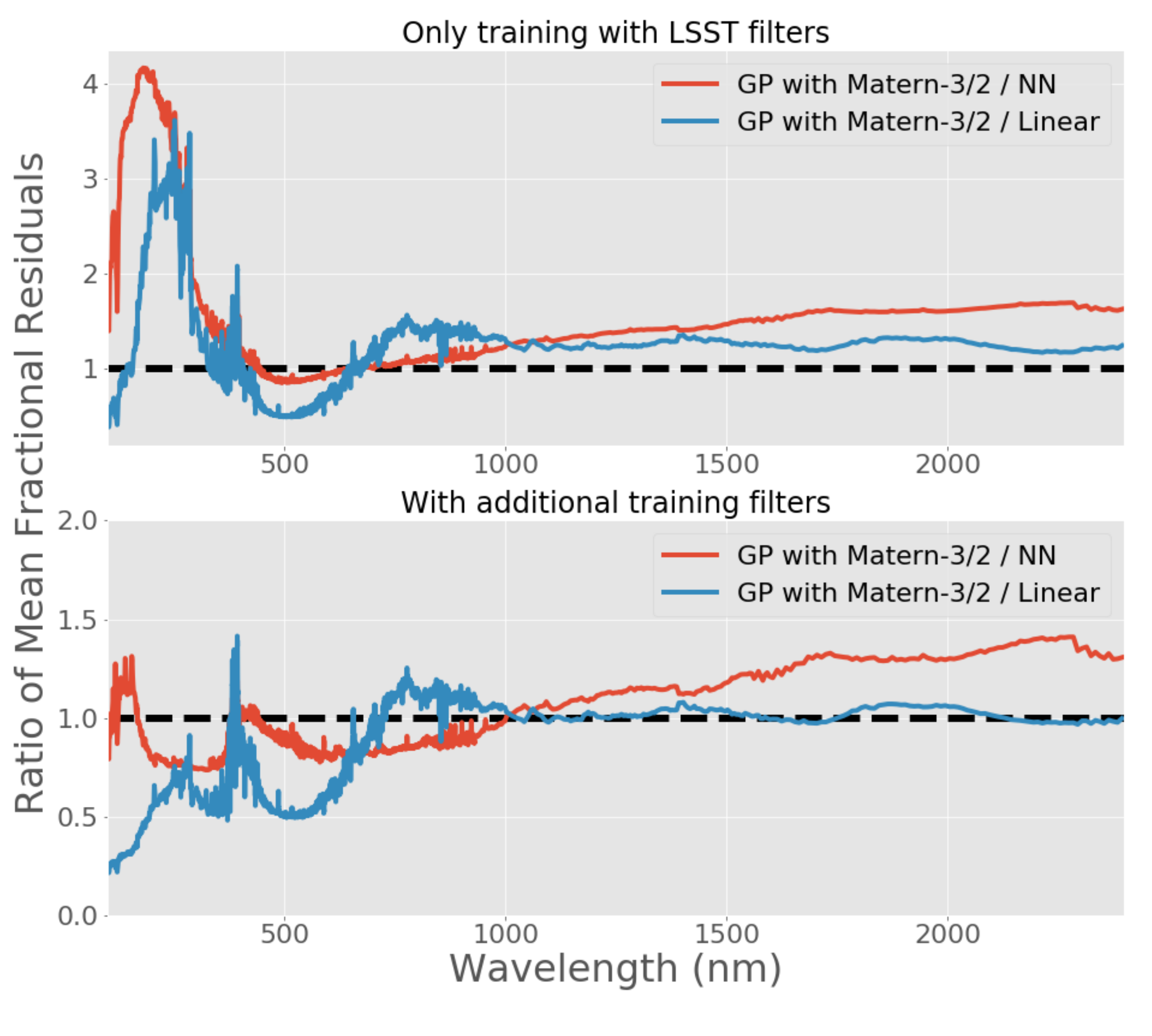}
	\caption{Top: Ratio of our method's fractional error to that from nearest neighbor and linear interpolation training only with 5 LSST colors.  Bottom: Ratio of the same methods using 9 colors to train the hyperparameters of the Gaussian Process.}	\label{spec_error_ratio_3}
\end{figure}

% \begin{figure}
%  	\centering
% 		\includegraphics[width=\linewidth]{fig_7.pdf}
% 	\caption{Top: Ratio of our method's fractional error to that from nearest neighbor and linear interpolation using spectra with wavelengths from 300 - 1200 $nm$.  Bottom: Ratio of the same methods comparing the interquartile mean of the residuals.}	\label{spec_error_ratio_4}
% \end{figure}

In general, there is an improvement in our estimates of the full spectrum compared to test 2. Figure \ref{spec_error_ratio_3} shows our results for the mean fractional residuals \replaced{as well as the comparison to the interquartile mean}{using the Matern-3/2 kernel with and without the artificial filters. There is obvious improvement across the whole spectral range, but most especially in the shorter wavelengths that now have training filters covering that wavelength range.} Table \ref{test_3_error} shows the new values for the results using the Gaussian process regression with new top hat filters. The mean residual error using the \replaced{exponential kernel decreased in the new test from 43.7\% across the whole spectrum in test 2 to 32.1\%. While it still does not match the 27.1\% of the nearest neighbor method, it beats the 2 nearest neighbor method and greatly outperforms the linear estimation which is up at 56.5\%.}{GP method decreased with every kernel. For the Matern-3/2 kernel it dropped from 51.2\% to 25.6\% now beating the nearest neighbor and linear techniques.} In the range between 300 and 1200 $nm$ there is improvement \replaced{for the exponential kernel which now is down to a mean error of 12.1\% while there is a slight increase for the squared exponential kernel to 21.6\%. The exponential kernel now outperforms all the other methods in the optical wavelength range as is clearly shown in Figure ??.}{for 3 of the 4 kernels as well.} Finally, we created an interquartile mean by only using the middle 50\% of estimated SEDs for each method based upon the mean error of the overall spectrum. We wanted to look at the interquartile mean in order to see if outliers were having a large effect on our estimated SEDs in this test. The bottom row of Table \ref{test_3_error} shows training with the additional filters leads our GP method with the exponential kernel to perform the best by a considerable amount according to this measurement. \added{This suggests that the exponential kernel does a good job reproducing the spectra, but also produces larger errors in some spectra compared to the Matern-3/2 kernel. Understanding where these errors are present could be possible since Gaussian Processes produce estimates of the variance in its results and this is addressed in our discussion of Future Work in \ref{discussion}. Knowing when to accept and when to reject our estimates could allow us to produce an even better set of estimated SEDs by combining estimates from multiple kernels.} 

Overall, the results of this test indicate that adding artificial filters helps our method \added{with all kernels} but we still need to do further work to reduce outliers and produce more accurate SEDs consistently. So far, we have just tested using simple top hat filters and additional development of these artificial filters can attempt to maximize the information used to identify the relationships in color space between spectra. \added{For now though it is encouraging that the simple additional filters lead the Matern-3/2 kernel to beat the nearest neighbor and linear methods when estimating a more complete spectrum and for the exponential kernel to beat them decisively when looking at the interquartile mean results.} Further work will hopefully help to reduce the outliers \replaced{in order to consistently outperform}{and lead to further improvements in the GP method over} the other methods across the full wavelength range.

\subsection{Extrapolation in Color Space} \label{extrapolation}

\begin{figure}
 	\centering
		\includegraphics[width=\linewidth]{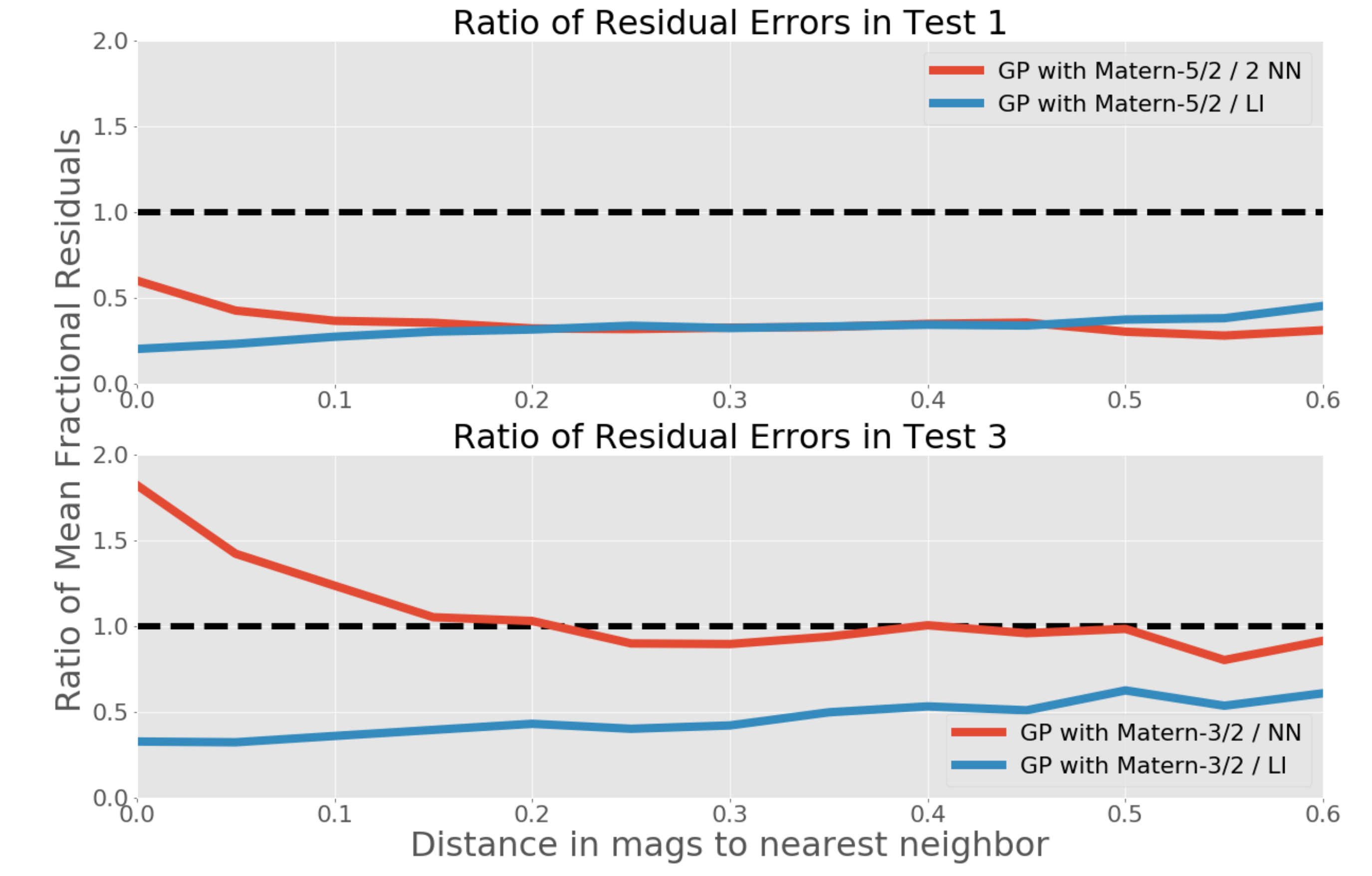}
	\caption{Here we compare the ratio of mean residual errors between SED estimation methods as a function of Euclidean distance to the nearest training set point. In both cases the Gaussian Process method improves results relative to the nearest neighbor as the distance to the nearest training point gets further. Top: Results from Section \ref{optical_test}. Bottom: Results from Section \ref{training_new_filters}}	\label{distance_error_ratio}.
\end{figure}

One of the challenges of using template sets is that available templates do not always cover the color space of interest. Improving estimates of SEDs in regions of color space that require extrapolation is one area that our method is able to improve compared to the other methods. In order to show this we sorted all the results from the previous tests by Euclidean distance from the color coordinate of the test point to the nearest training neighbor. Figure \ref{distance_error_ratio} shows the results for all objects within a radius of \replaced{0.4}{0.6} magnitudes in color space to a training point. This covers \replaced{almost 90\%}{over 95\%} of our test points. The top plot in Figure \ref{distance_error_ratio} shows the results from Section \ref{optical_test} where the GP method outperforms all the way through these distances but achieves best results over the nearest neighbors at larger distances going to lower than one-third of the error in the nearest neighbor and linear interpolation methods \replaced{at distances over .2 mags away from the nearest training data}{consistently across the distances}. In the bottom plot we show results from Section \ref{training_new_filters}. In that test we saw that overall the nearest neighbors method was able to beat our GP method, but looking at this plot we see that the GP method is able to catch up \replaced{to}{and beat} the nearest neighbors method when the test colors get further away from the color space of the original template set. \deleted{Beyond a certain point the nearest neighbor begins to beat our method again as the PCA coefficients from our GP start to get far enough away from the training data that they approach the mean value of the GP which we set at 0.} These results show that our GP estimation method is a powerful new tool for extrapolating in color space. 

\subsection{Using Narrowband Filters} \label{narrowband}

\added{In order to further show the capabilities of our technique we also applied it to a set of narrowband filters once again using the same SED sets as we did in \ref{estimate_seds}. We chose a set of 4 narrowband filters from the Hubble Space Telescope ACS/HRC \footnote{Filter transmission curves collected from: \url{http://svo2.cab.inta-csic.es/svo/theory/fps3/index.php?mode=browse&gname=HST&gname2=ACS_HRC}}. The filters are centered at approximately 344, 502, 658 and 892 nm and range in effective width from 56.7 to 149.1 nm. We ran the same tests as \ref{optical_test} and \ref{training_new_filters} and present the results in Tables \ref{test_4_4_1_results} and \ref{test_4_4_3_results}.

The results are very similar to the wider LSST filters. In the first test we outperform the nearest neighbor and linear methods in 3 of the 4 kernels with mean residual error as low as 3.78\% for the Matern-3/2 kernel. We also tested on spectra extended to 99-2400 $nm$ using the same artificial filters plus the LSST \textit{y} filter as training filters to fit the hyperparameters like we did in \ref{training_new_filters}. We added the LSST \textit{y} filter to the training set to bridge the gap between the reddest narrowband filter and the red top hat filters. In this test we get comparable results to the nearest neighbors method when using the Matern-3/2 kernel. Once again, a refined set of training filters may be able to further improve our results, but for a simple set the results are encouraging that even with the limited amount of the spectrum sampled by broadband filters and with only 4 filters we are able to estimate a large portion of the spectrum to almost the same accuracy.}

\begin{deluxetable}{cc|ccccccc}
\tablecaption{Percentage residual errors in flux with narrowband filters limiting SEDs to 299-1200 nm.}
%\hline
\tablehead{
\colhead{Error Metric} & \colhead{Wavelength Range} & \colhead{Exp.} & \colhead{Sq. Exp.} & \colhead{Mat.-3/2} & \colhead{Mat.-5/2} & \colhead{NN} & \colhead{2 NN} & \colhead{Linear} \\
\colhead{} & \colhead{nm} & \colhead{} & \colhead{} & \colhead{} & \colhead{} & \colhead{} & \colhead{} & \colhead{}
}
\startdata
Mean Error & 299 - 1200 & 7.51\% & 4.74\% & 3.78\% & 3.98\% & 10.1\% & 9.57\% & 7.42\%  \\
Interquartile Mean Error & 299 - 1200 & 4.38\% & 2.71\% & 2.80\% & 2.53\% & 7.74\% & 6.53\% & 5.17\% \\
\enddata
%\end{tabular}%
%}
\tablecomments{Residuals are errors in flux between true and estimated SEDs.} \label{test_4_4_1_results}
\end{deluxetable}

\begin{deluxetable}{cc|ccccccc}
%\hline
\tablecaption{Percentage Residual Errors in flux for with narrowband filters extending SEDs to 99-2400 nm.}
\tablehead{
\colhead{Error Metric} & \colhead{Wavelength Range} & \colhead{Exp.} & \colhead{Sq. Exp.} & \colhead{Mat.-3/2} & \colhead{Mat.-5/2} & \colhead{NN} & \colhead{2 NN} & \colhead{Linear} \\
\colhead{} & \colhead{nm} & \colhead{} & \colhead{} & \colhead{} & \colhead{} & \colhead{} & \colhead{} & \colhead{}
}
\startdata
Mean Error & 99 - 2400 & 31.0\% & 137.7\% & 27.6\% & 58.4\% & 25.4\%  & 30.6\% & 33.6\% \\
Mean Error & 299 - 1200 & 12.5\% & 33.6\% & 14.9\% & 21.7\% & 15.1\% & 14.3\% & 11.9\% \\
Interquartile Mean Error & 99 - 2400 & 11.5\% & 21.8\% & 16.6\% & 15.9\% & 13.8\% & 13.1\% & 12.1\%  \\
\enddata
%\end{tabular}
%}
\tablecomments{Residuals are errors in flux between true and estimated SEDs.} \label{test_4_4_3_results}
\end{deluxetable}

\section{Expanding Template Sets for Photometric Redshifts} \label{photoZ}

While the overall goal of our estimation method is to expand template sets for a variety of applications, one of the most common uses for template sets is in estimating photometric redshifts. As a demonstration of the benefits of our technique we performed photometric redshift estimation on a mock catalog with BC03 training spectra. In this section, we first establish the benefits of larger template sets on photometric redshift estimation. We then show how our technique can be used to improve photometric redshift estimation by expanding a template set through the generation of new estimated SEDs at specific locations in color space. Additionally, it is important to note that while Gaussian Processes have been applied to photometric redshifts previously\replaced{ by \citet{gpz} and \citet{gpphotoz}}{\citep{gpphotoz_sdss, gpz, gpphotoz}} our method is a general framework for expanding template sets and constructing SEDs with PCA coefficients. Below we use an existing photometric redshift method with an expanded template set as an example application since it is easy to compare results by plugging in different template sets with the same catalogs and code. Other possible applications of our method include representing large sets of SEDs with only a few PCA coefficients or creating continuous distributions of SEDs using features other than colors. We discuss these further in Section \ref{discussion}.

\subsection{Mock Catalog and Method}
Our simulated catalog was 100,000 objects from a larger catalog used in \citep{lsstphotoz} created using the galaxy formation model of \citet{gpcatalog} with redshifts up to z=6, but we focus our results on the 89,471 objects with z$\leq$3. The photometry in the catalog is in the SDSS ugriz and Pan-Starrs y filters. The catalog came without errors so we generated photometric errors of a simulated LSST-like survey using the LSST simulations software stack \citep{sims}. 

To calculate redshifts we used LePHARE (PHotometric Analysis for Redshift Estimations)\footnote{http://www.cfht.hawaii.edu/$\sim$arnouts/LEPHARE/lephare.html} \citep{lephare1, lephare2}. The code is an SED template fitting code that uses a chi-square method comparing the photometric flux of the templates at a sequence of redshifts to the catalog photometry to find the best template and redshift match for each galaxy.

\subsection{Expanding template sets} \label{template_sets}

We wanted to consider the impact of the number of templates used in a photometric redshift analysis. Along with the original set of 10 template SEDs, we created a set of 60 BC03 templates by randomly selecting 50 additional spectra from the full SED library and a set of the 10 original templates plus 50 templates created using an exponential kernel function trained with the method explained in Section \ref{training_new_filters}. We used the SDSS ugriz and Pan-Starrs y filters along with the same 4 top-hat filters from before. To decide where in color space to estimate new colors we performed a k-means clustering on the catalog in the SDSS ugriz + Pan-Starrs y color space with k=50. We then used the locations of the color centers as the locations to estimate new SEDs. While this method of estimating color locations is done on the redshifted colors and thus will cover some areas with unrealistic rest frame colors it will make sure the rest frame SEDs of the majority of galaxies are covered. We then estimated redshifts using LePHARE for all objects in the mock catalog using the 3 template sets. \added{Since we were trying to isolate the effect of changing template sets on the redshift estimation we included no priors when running the code. Therefore, the results presented here are a worst case scenario run without optimizing the redshift estimation in other ways.} Table \ref{template_compare_table} shows the results of four statistics calculated from the photo-z estimation. We define the distance corrected difference in redshift to be $\Delta z = \frac{z_{true} - z_{phot}}{1 + z_{true}}$ and calculate four statistics as used by the LSST:
\begin{itemize}
\item Bias: Outlier rejected bias. Mean of differences between true and estimated redshift across interquartile range (IQR). $\overline{\Delta z_{25-75\%}}$
\item Standard Deviation: Standard deviation of $\Delta z$. $\sqrt[]{\overline{(\Delta z - \overline{\Delta z})^{2}}}$
\item Standard Deviation of Interquartile Range: The spread of the difference between true and estimated redshift for the middle 50\% of differences divided by 1.349 to compare to standard deviation. $\frac{\Delta z_{75\%} - \Delta z_{25\%}}{1.349}$
\item Fraction of Outliers: The fraction of catalog objects with a difference between true and estimated redshift greater than 0.06 or 3 times the Standard Deviation of the IQR whichever is greater.
\end{itemize}

%\begin{table}
%\centering
%\resizebox{\columnwidth}{!}{%
\begin{deluxetable}{c|cccc}
%\hline
\tablecaption{Statistics for photometric redshifts estimated for catalog objects with z $\leq$ 3 \label{template_compare_table}} 
\tablehead{
Template Set & Bias & Std. Dev. & IQR St. Dev. & Outlier Frac.
}
\startdata
10 BC03 Templates & 0.051 & 0.279 & 0.121 & 0.196 \\ 
60 BC03 Templates &  0.013 & 0.173 & 0.036 & 0.154 \\ 
10 BC03 + 50 Exp Kernel & 0.007 & 0.217 & 0.048 & 0.131 \\
\enddata
%\end{tabular}%
\end{deluxetable}

\begin{figure}
 	\centering
		\includegraphics[width=\linewidth]{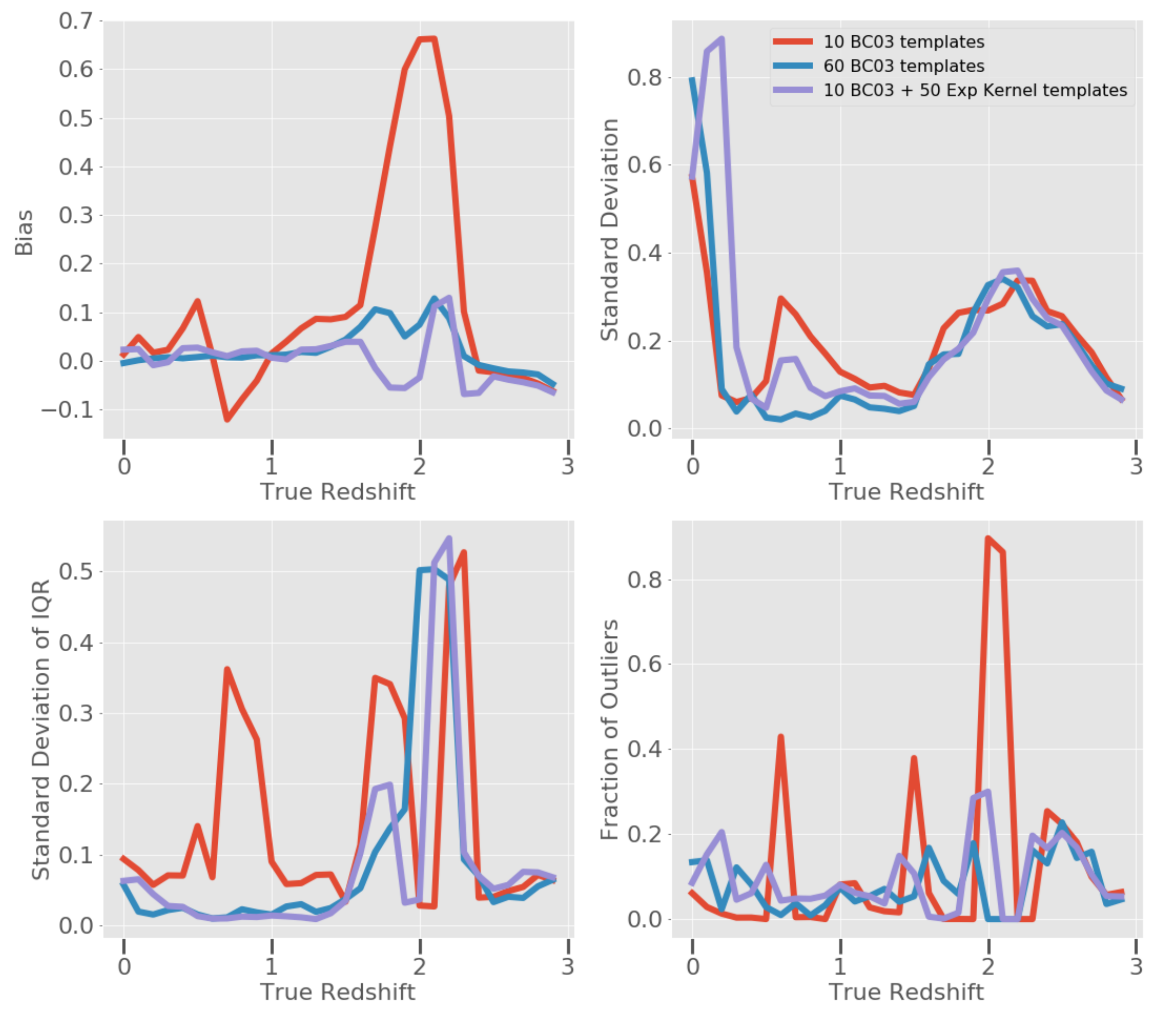}
	\caption{Comparing descriptive statistics from photometric redshift estimation with 10 and 60 BC03 templates to 10 BC03 templates + 50 estimated SEDs created using the Gaussian Process estimation method with an exponential kernel in color space. The addition of Gaussian process interpolated templates improves the redshift estimation in the range z $\leq$ 3 compared to the original 10 templates from which they are derived and produce results comparable to using 60 BC03 templates.}	\label{interp_redshift_comparison}
\end{figure}

Table \ref{template_compare_table} and Figure \ref{interp_redshift_comparison} show that using a larger template set improves the redshift estimation in the range z $\leq$ 3. The standard deviation for the redshift residuals is reduced by 38\% when going from 10 to 60 and the bias falls by 74\%. Therefore, the coverage of templates in color space does make a difference in the accuracy of photometric redshift estimation and increasing the number of templates is beneficial. While this is an idealized case since we used the templates that were used to create the simulated catalog colors, we do expect to see that using our method to create additional realistic template SEDs will provide measurable improvement in photometric redshift estimation.

\begin{figure}
 	\centering
		\includegraphics[width=\linewidth]{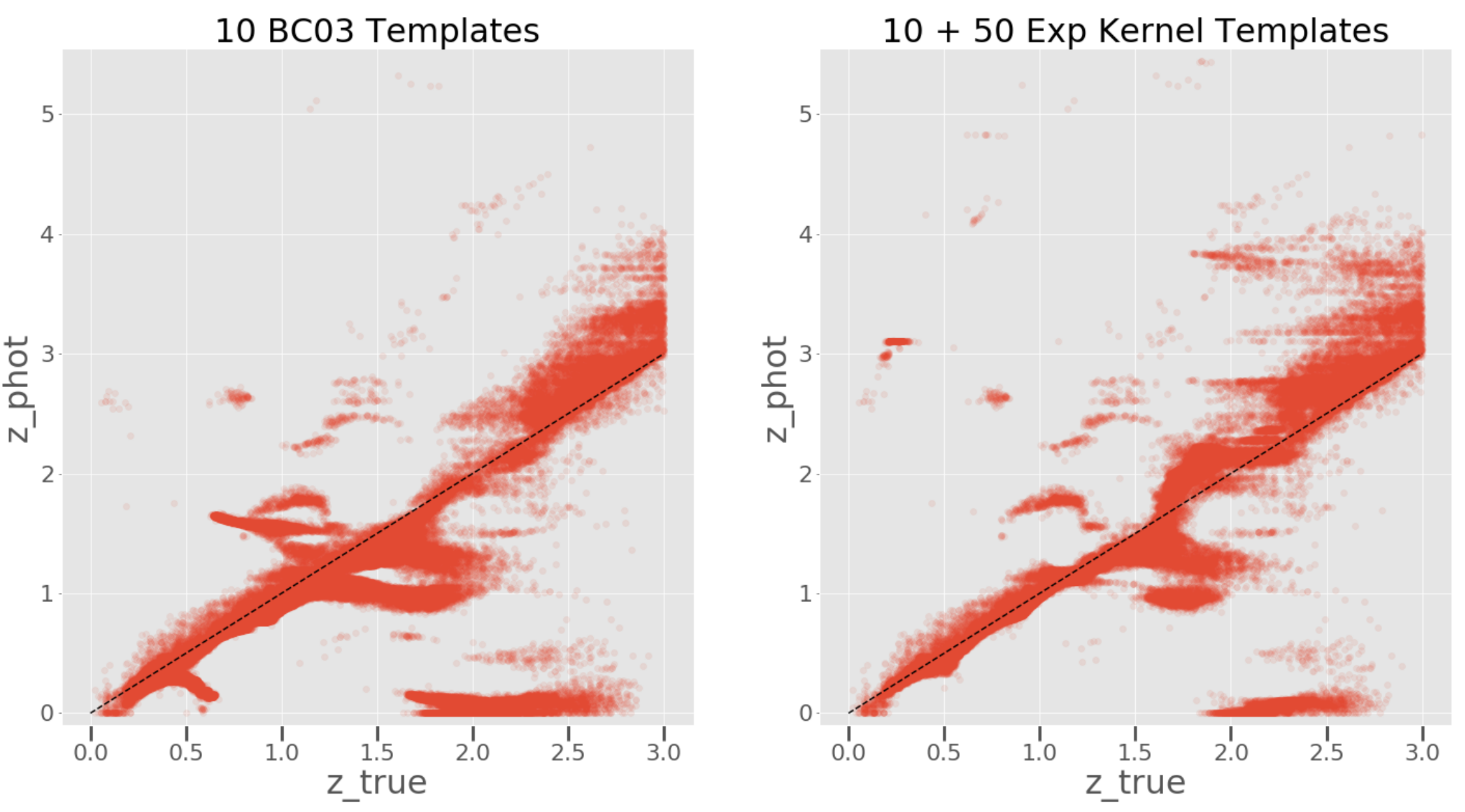}
	\caption{Scatter plots comparing the true redshift from the mock catalog and the estimated redshift from photometry when using 10 BC03 templates (left) and adding 50 estimated SEDs using our technique with an exponential kernel (right). Notice how the additional templates help eliminate some of the horizontal features that appear when only using the 10 templates on their own.}	\label{photoz_scatter}
\end{figure}

Comparing the exponential kernel method to the original 10 templates reveals significant improvements when using our technique to expand the template set.  Our estimated SEDs successfully improve estimates across all measured statistics in the range z $\leq$ 3 including improving the standard deviation by 22.0\%, the standard deviation of the IQR by 60.6\% and the bias by 86.2\%. \added{Figure \ref{photoz_scatter} compares the scatter plots for the two runs side-by-side. The exponential kernel templates help eliminate some of the longer horizontal features found in the scatter plot for the 10 templates on their own.} 

We can also consider our 50 estimated SEDs against the results from adding 50 of the original template SEDs. The 60 template set in Section \ref{template_sets} had a better standard deviation in its estimates 0.173 to our 0.217 and the standard deviation of the IQR at 0.036 to our 0.048, but we were able to better it in the bias by 0.006. Furthermore, as noted above the 60 templates were an idealized case and it is not surprising that we were not able to match the standard deviation. But, it is a very positive sign that our best 50\% of results represented in the IQR standard deviation and outlier rejected bias approach the levels of the 60 BC03 templates and in the case of the bias are able to improve upon that set.

\section{Future Work} \label{discussion}

We have presented the results of our estimation technique to produce realistic and useful SEDs for studying galaxies. While it is best at reproducing the flux in the wavelength range of the filters it uses we have also shown we can improve upon existing techniques for estimating spectra even outside the range of the colors we have for a galaxy. This already improves photometric redshift estimation by reducing the standard deviation in the residuals by 24.5\% compared to a base template set as shown in Section \ref{photoZ}. There were also improvements to the bias and the standard deviation of the IQR. Further work can be done to improve the artificial filters for greater accuracy at wavelengths beyond the optical range. Our simple top hat filters were only a first effort at solving this issue and we will look at how to design artificial bandpasses that maximize using the information in wavelengths only available to the training set.

Additionally, Gaussian process regression provides a measure of the variance around the mean estimate for each input value. In our method we only used the mean value for eigencoefficients from the GP to create new SEDs, but there is information in the variance results of the GP that we could use to quantify the uncertainty in our predicted SEDs. This uncertainty could help in deciding which areas of color space can be confidently extrapolated with our technique using a given training set. \added{Or as mentioned in \ref{training_new_filters} we could use the information to combine the best estimates from a set of kernels to cover color space more completely and accurately than only using a single kernel.} In Section \ref{extrapolation} we showed our method is better than others at extrapolating to new areas of color space, but also has limits due to the nature of the Gaussian Processes. Understanding the limits may come from studying the accompanying error estimates. We may also use this information to provide uncertainty estimates on our predicted SEDs. We could then use the information in Bayesian analyses like those used in photometric redshift estimation.

Finally, we want to extend the use of the tools developed in this paper to other applications. The ability to describe a basis for constructing SEDs with PCA coefficients will allow galaxy evolution codes to retain a galaxy's SED at each time step in a simulation. Storing only a set of around 10 PCA coefficients will allow this to be practical in terms of memory use compared to storing the full spectrum at each time step. Furthermore, we could explore creating continuous distributions of galaxy SEDs in other feature spaces such as metallicity and age. This would help semi-analytical models of galaxies create more realistic color distributions in mock catalogs by providing SEDs that are not restricted to the finite grid of metallicity, age and other properties that simulated SEDs like BC03 currently allow.

\section{Conclusion} \label{conclusion}

We have shown that Gaussian process interpolation of template set eigencoefficients is able to create a continuous interpolation and extrapolation from a training set of SEDs to the SEDs for other points in color space. This mapping provides SEDs that improve upon standard interpolation techniques currently used both in the estimation of the true spectrum and in the generation of colors from the predicted SEDs. For the wavelength range where photometric filters and spectra overlap we can improve the mean error estimate of the spectrum for a given location in color space by over 65\%. Furthermore, Section \ref{extrapolation} indicates that the best improvements come as the test points get further from the training data. As an example application we demonstrated that our method can help photometric redshift estimation. We improved the standard deviation of the error in photometric redshifts by over 24.8\% and lowered the outlier rejected bias by over 87.5\% compared to the original template set. In the future we hope to extend the applications to improving outputs for galaxy evolution modelling codes and other areas where SEDs are used to calculate galaxy properties or in generation of mock catalogs. Overall, this technique is a powerful addition to the astronomical toolbox anywhere interpolation or extrapolation of template SEDs would be useful and our Python code, ESP (\textbf{E}stimating \textbf{S}pectra from \textbf{P}hotometry), is documented and openly available on github at \url{https://github.com/jbkalmbach/esp}. It also includes a jupyter notebook with the code to reproduce all plots in our paper.
\vspace{\baselineskip}
%\section*{Acknowledgments}

The authors would like to thank Chris Morrison and Scott Daniel for reading and providing comments on early drafts of this manuscript. This work was supported by the U.S. Department of Energy, Office of Science, under Award Number DE-SC-0011635.

\software{ESP \citep{esp}, Numpy \citep{numpy}, Matplotlib \citep{matplotlib}, Scipy \citep{scipy}, Scikit-Learn \citep{scikit}, LSST Simulations Software Stack \citep{sims}, George \citep{george}, Lephare \citep{lephare1, lephare2}, Pandas \citep{pandas}, Jupyter \citep{jupyter}}

{}

\end{document}